\documentclass[prl,aps,twocolumn,showpacs,nofootinbib]{revtex4}

\usepackage{graphicx}
\usepackage{stmaryrd}
\usepackage{rotating}
\usepackage{amsmath}
\usepackage{amsfonts}
\usepackage{amssymb}
\usepackage[countmax]{subfloat}
\usepackage{dcolumn} 
\usepackage{bm} 
\usepackage{color}

\usepackage{float}
\usepackage{latexsym}

\begin{document}

\title{Dissipationless Multiferroic Magnonics}



\author{Wei Chen$^{1}$ and Manfred Sigrist$^{2}$} 

\affiliation{$^{1}$Max-Planck-Institut f$\ddot{u}$r Festk$\ddot{o}$rperforschung, Heisenbergstrasse 1, D-70569 Stuttgart, Germany 
\\
$^{2}$Theoretische Physik, ETH-Z\"urich, CH-8093 Z\"urich, Switzerland}

\date{\rm\today}

\begin{abstract}

{We propose that the magnetoelectric effect in multiferroic insulators with coplanar antiferromagnetic spiral order, such as BiFeO$_{3}$, enables electrically controlled magnonics without the need of a magnetic field. Applying an oscillating electric field in these materials with frequency as low as household frequency can activate Goldstone modes that manifests fast planar rotations of spins, whose motion is essentially unaffected by crystalline anisotropy. Combining with spin ejection mechanisms, such a fast planar rotation can deliver electricity at room temperature over a distance of the magnetic domain, which is free from energy loss due to Gilbert damping in an impurity-free sample.   }

\end{abstract}





\maketitle

{\it Introduction.-} A primary goal of spintronic research is to seek for mechanisms that enable electric (${\bf E}$) field controlled spin dynamics, since, in practice, ${\bf E}$ fields are much easier to manipulate than magnetic (${\bf B}$) fields. As spins do not directly couple to ${\bf E}$ field, incorporating spin-orbit coupling seems unavoidable for this purpose. Along this line came the landmark proposals such as spin field effect transistor \cite{Datta90} and spin-orbit torque \cite{Manchon08,Pesin12,Haney10,Haney13}, the realizations of which suggest the possibility of spin dynamics with low power consumption. On the other hand, in another major category of spintronics, namely magnonics, which aims at the generation, propagation, and detection of magnons, a mechanism that enables electrically controlled magnonics without the aid of a magnetic field has yet been proposed. 

Raman scattering experiments \cite{Rovillain09,Rovillain10} on the room temperature multiferroic BiFeO$_{3}$ (BFO) shed light on this issue. The magnetic order of BFO is a canted antiferromagnetic (AF) spiral on the plane spanned by the electric polarization ${\bf P}$ along $\left[111\right]$ and one of the three symmetry-equivalent wave vectors on a rhombohedral lattice  \cite{Catalan09,Lebeugle08}. The spins have only a very small out-of-plane component \cite{Ederer05,Ramazanoglu11}.  Applying a static ${\bf E}$ field $\sim 100$kV/cm significantly changes the cyclon (in-plane) and extra-cyclon (out-of-plane) magnons because of the magnetoelectric effect \cite{Rovillain10}. Indeed, spin-orbit coupling induced magnetoelectric effects   are a natural way to connect ${\bf E}$ field to the spin dynamics of insulators \cite{Moriya60,Katsura05}. Motivated by the Raman scattering experiments on BFO, in this Letter we propose that applying an oscillating ${\bf E}$ field to a coplanar multiferroic insulator (CMI) that has AF spiral order can achieve electrically controlled dissipationless magnonics, which can deliver electricity with frequency as low as household frequency up to the range of magnetic domains. Compared to the magnonics that uses ${\bf B}$ field, microwave, or spin torques to generate spin dynamics in prototype Y$_{3}$Fe$_{5}$O$_{12}$ (YIG) \cite{Kruglyak10,Serga10,Kajiwara10}, the advantage of using CMI is that a single domain sample up to mm size is available \cite{Johnson13}, and Raman scattering data indicate well-defined magnons in the absence of ${\bf B}$ field \cite{Rovillain10}, so an external ${\bf B}$ field is not required in the proposed mechanism.

\begin{figure}[ht]
\begin{center}
\includegraphics[clip=true,width=0.95\columnwidth]{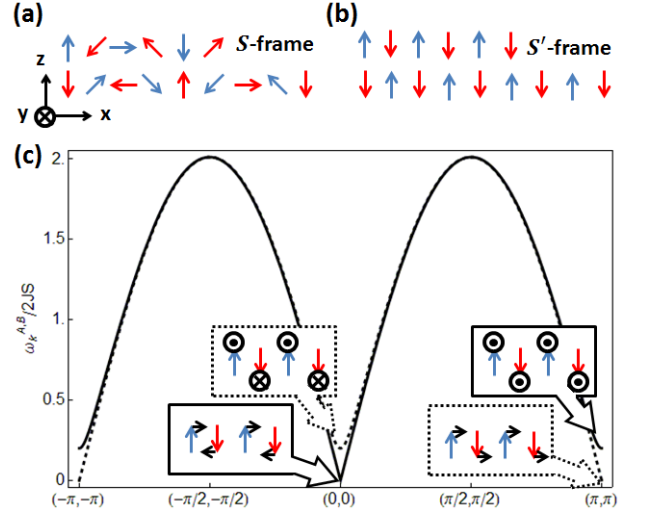}
\caption{ (color online) Schematics of 2D AF spiral in the (a) original $S$-frame and the (b) rotated $S^{\prime}$-frame. Red and blue arrows indicate the spins on the two sublattices. (c) Spin wave dispersion $\omega_{\bf k}^{+}$ (dashed line) and $\omega_{\bf k}^{-}$ (solid line) solved in the $S^{\prime}$ frame, with $D_{a}/J=0.14$, $D_{c}=0$. Inserts show their eigen modes in the $S^{\prime}$ frame near ${\bf k}=(0,0)$ and $(\pi,\pi)$, where the spin dynamics $d{\bf S}_{i}^{\prime}/dt$ is indicated by black arrows or symbols.} 
\label{fig:AF_spiral_S_Sp_frame}
\end{center}
\end{figure}

{\it Spin dynamics in CMI.-} We start from the AF spiral on a square lattice shown in Fig.~\ref{fig:AF_spiral_S_Sp_frame} (a), described by
\begin{eqnarray}
H=\sum_{i,\alpha}J{\bf S}_{i}\cdot{\bf S}_{i+\alpha}-{\bf D}_{\alpha}\cdot\left({\bf S}_{i}\times{\bf S}_{i+\alpha}\right)
\label{AF_DM_Hamiltonian}
\end{eqnarray}
where ${\boldsymbol\alpha}=\left\{{\bf a},{\bf c}\right\}$ are the unit vectors defined on the $xz$-plane, $J>0$, and ${\bf D}_{\alpha}=D_{\alpha}{\bf {\hat y}}>0$ is the Dzyaloshinskii-Moriya (DM) interaction. The staggered moment $(-1)^{i}{\bf S}_{i}$ in the ground state shown in Fig.~\ref{fig:AF_spiral_S_Sp_frame} (a) is characterized by the angle $\theta_{\alpha}={\bf Q}\cdot{\boldsymbol\alpha}=-\sin^{-1}\left(D_{\alpha}/\tilde{J}_{\alpha}\right)$ between neighboring spins, where $\tilde{J}_{\alpha}=\sqrt{J^{2}+D_{\alpha}^{2}}$. The DM interaction 
\begin{eqnarray}
{\bf D}_{\alpha}={\bf D}_{\alpha}^{0}+{\overline w}{\bf E}\times{\boldsymbol \alpha}
\label{DM_E_field}
\end{eqnarray}
can be controlled by an ${\bf E}$ field \cite{Shiratori80}, where ${\bf D}_{\alpha}^{0}$ represents the intrinsic value due to the lack of in version symmetry of the $\alpha$-bond. In the rotated reference frame $S^{\prime}$ defined by 
\begin{eqnarray}
S_{i}^{\prime z}&=&S_{i}^{z}\cos{\bf Q}\cdot{\bf r}_{i}+S_{i}^{x}\sin{\bf Q}\cdot{\bf r}_{i}\;,
\nonumber \\
S_{i}^{\prime x}&=&-S_{i}^{z}\sin{\bf Q}\cdot{\bf r}_{i}+S_{i}^{x}\cos{\bf Q}\cdot{\bf r}_{i}\;,
\label{frame_rotation}
\end{eqnarray}
and $S_{i}^{\prime y}=S_{i}^{y}$, the Hamiltonian is 
\begin{eqnarray}
H&=&\sum_{i,\alpha}\tilde{J}_{\alpha}\left(S_{i}^{\prime x}S_{i+\alpha}^{\prime x}+S_{i}^{\prime z}S_{i+\alpha}^{\prime z}\right)+JS_{i}^{\prime y}S_{i+\alpha}^{\prime y}\;.
\label{AF_DM_Hamiltonian_rotated}
\end{eqnarray}
Since $\tilde{J}_{\alpha}>J$, the spins have collinear AF order and all $S_{i}^{\prime z}=(-1)^{i}S$ lie in $xz$-plane.

The spin dynamics in the absence of ${\bf B}$ field is governed by the Landau-Lifshitz-Gilbert (LLG) equation
\begin{eqnarray}
\frac{d{\bf S}_{i}^{\prime}}{dt}=\frac{\partial H}{\partial{\bf S}_{i}^{\prime}}\times{\bf S}_{i}^{\prime}+\alpha_{G}{\bf S}_{i}^{\prime}\times\left(\frac{\partial H}{\partial{\bf S}_{i}^{\prime}}\times{\bf S}_{i}^{\prime}\right)
\label{EOM_with_damping_Sp_frame}
\end{eqnarray}
expressed in the $S^{\prime}$ frame, where $\alpha_{G}$ is the phenomenological damping parameter. Eq.~(\ref{EOM_with_damping_Sp_frame}) can be solved by the spin wave ansatz for the even ($e$) and odd ($o$) sites \cite{deSousa08}
\begin{eqnarray}
\left(
\begin{array}{l}
S_{e,o}^{\prime x} \\
S_{e,o}^{\prime y} 
\end{array}
\right)
=
\left(
\begin{array}{l}
u_{e,o}^{x} \\
v_{e,o}^{y} 
\end{array}
\right)e^{i\left({\bf k}\cdot{\bf r}_{e,o}-\omega t\right)}\;.
\label{spin_wave_ansatz}
\end{eqnarray}
Ignoring the damping term in Eq.~(\ref{EOM_with_damping_Sp_frame}) yields eigenenergies 
\begin{eqnarray}
\frac{\omega_{\bf k}^{\pm}}{2S}&=&\left[\left(\sum_{\alpha}\tilde{J}_{\alpha}\pm\gamma_{\alpha-}({\bf k})\right)^{2}-\left(\sum_{\alpha}\gamma_{\alpha+}({\bf k})\right)^{2}\right]^{1/2}\;,
\label{AF_spiral_eigen_modes}
\end{eqnarray}
where $\gamma_{\alpha\pm}({\bf k})=\left(\tilde{J}_{\alpha}/2\pm J/2\right)\cos{\bf k}\cdot{\boldsymbol\alpha}$. Their eigenvalues and eigenvectors near ${\bf k}=(0,0)$ and ${\bf k}=(\pi,\pi)$ are summarized below
\begin{eqnarray}
&&\left\{\omega_{{\bf k}\rightarrow (0,0)}^{+},\omega_{{\bf k}\rightarrow(\pi,\pi)}^{-}\right\}=2S\sqrt{2(D_{a}^{2}+D_{c}^{2})}\;,
\nonumber \\
&&\left(
\begin{array}{l}
u_{e} \\
v_{e} \\
u_{o} \\
v_{o}
\end{array}
\right)\propto\left(
\begin{array}{l}
0 \\
1 \\
0 \\
\mp 1
\end{array}
\right)+{\cal O}\left(\frac{D}{J}\right)\;.
\nonumber \\
&&\left\{\omega_{{\bf k}\rightarrow (0,0)}^{-},\omega_{{\bf k}\rightarrow(\pi,\pi)}^{+}\right\}=0\;,\;\left(
\begin{array}{l}
u_{e} \\
v_{e} \\
u_{o} \\
v_{o}
\end{array}
\right)\propto\left(
\begin{array}{l}
1 \\
0 \\
\mp 1 \\
0
\end{array}
\right)\;.
\label{Goldstone_mode_solution}
\end{eqnarray}
The in-plane magnon $d{\bf S}_{i}^{\prime}/dt=(dS_{i}^{\prime x}/dt,0,0)$ is gapless, while the out-of-plane magnon $d{\bf S}_{i}^{\prime}/dt=(0,dS_{i}^{\prime y}/dt,0)$ develops a gap, as displayed in Fig.~\ref{fig:AF_spiral_S_Sp_frame} (c). Even including the damping term in Eq.~(\ref{EOM_with_damping_Sp_frame}), the in-plane magnons very near the Goldstone modes $\omega_{{\bf k}\rightarrow(0,0)}^{-}$ and $\omega_{{\bf k}\rightarrow(\pi,\pi)}^{+}$ remain unchanged and damping-free.  
Away from the Goldstone limit, the eigenenergies become complex, hence the magnons are subject to the damping and decay within a time scale set by $\alpha_{G}^{-1}$.

\begin{figure}[ht]
\begin{center}
\includegraphics[clip=true,width=0.95\columnwidth]{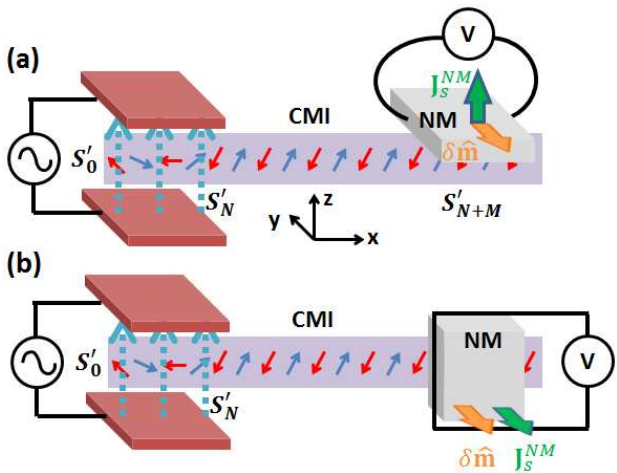}
\caption{ (color online) Experimental proposal of using oscillating ${\bf E}$ field to induce spin dynamics in CMI. The AF spiral order is shown in the $S^{\prime}$ frame. The ${\bf E}$ field is applied between ${\bf S}_{0}^{\prime}$ and ${\bf S}_{N}^{\prime}$, causing dynamics in the whole spin texture. Two ways for spin ejection out of ${\bf S}_{N+M}^{\prime}$ are proposed: (a) Using SHE to converted it into a charge current. (b) Using time-varying spin accumulation and inductance.  } 
\label{fig:E_field_spiral}
\end{center}
\end{figure}

{\it Spin dynamics induced by oscillating ${\bf E}$ field.-} We analyze now the spin dynamics in the damping-free in-plane magnon channel induced by magnetoelectric effects (Eq.~(\ref{Goldstone_mode_solution})). Unlike the spin injection by using the spin Hall effect (SHE) to overcome the damping torque \cite{Kajiwara10}, our design does not require an external ${\bf B}$ field, and is feasible over a broad range of frequencies. Consider the device shown in Fig.~\ref{fig:E_field_spiral}, where an oscillating electric field ${\bf E}={\bf E}^{0}\cos\omega t$ is applied parallel to the ferroelectric moment over a region of length $L=Na$, such that the DM interaction in Eq.~(\ref{DM_E_field}) oscillates in this region. Thus, the wave length of the spiral changes with time yielding an oscillation of the number of spirals inside this region,
\begin{eqnarray}
n_{Q} =\frac{L}{2\pi/|{\bf Q}|} \approx\frac{N}{2\pi J}\left[D_{a}^{0}+{\overline w}E^{0}a\cos\omega t\right]\;,
\end{eqnarray}
assuming $D_{a}=D_{a}^{0}+{\overline w}E^{0}a\ll J$, $D_{c}=0$, and ${\bf E}\perp{\bf a}$. Suppose the spin ${\bf S}_{0}$ at one boundary is fixed by, for instance, surface anisotropy because of specific coating. Then ${\bf S}_{N}$ at the other boundary rotates by 
\begin{eqnarray}
\frac{\partial\theta_{N}}{\partial t}=-\frac{N}{J}{\overline w}E^{0}a\omega\sin\omega t\;,
\label{dtheta_dt}
\end{eqnarray}
because whenever the number of waves $n_{Q}$ changes by $1$, ${\bf S}_{N}$ rotates $2\pi$ in order to to wind or unwind the spin texture in the ${\bf E}$ field region. The significance of this mechanism is that although the ${\bf E}$ field is driven by a very small frequency $\omega$, the spin dynamics $\partial_{t}\theta_{N}$ at the boundary is many orders of magnitude enhanced because of the winding process. The rotation of ${\bf S}_{N}$ serves as a driving force for the spin dynamics in the field-free region from ${\bf S}_{N}$ to ${\bf S}_{N+M}$. As long as the spin dynamics is slower than the energy scale of the DM interaction $\partial_{t}\theta_{i}<|{\bf D}_{0}|/\hbar\sim$THz, one can safely consider the ${\bf E}$ field region as adiabatically changing its wave length but remaining in the ground state. The spins in the field-free region rotate coherently $\partial_{t}\theta_{N}=\partial_{t}\theta_{N+1}=...=\partial_{t}\theta_{N+M}$, synonymous to exciting the $\omega_{{\bf k}
 \rightarrow(0,0)}^{-}$ mode in Eq.(\ref{Goldstone_mode_solution}), hence the spin dynamics in the field-free region remains damping-free in an ideal situation.

In real materials, crystalline anisotropy and impurities are the two major sources to spoil the spin rotational symmetry implicitly assumed here. In the supplementary material\cite{supplementary}, their effects are discussed by drawing analogy with similar situations in the atom absorption on periodic substrates and the impurity pinning of charge density wave states. It is found that crystalline anisotropy remains idle because of the long spiral wave length and the smallness of crystalline anisotropy compared to exchange coupling. The impurities that tend to pin the spins along certain crystalline direction open up a gap in the Goldstone mode and cause energy dissipation, which nevertheless do not obstruct the coherent rotation of spins generated by Eq.~(\ref{dtheta_dt}).

\begin{table}
\begin{tabular}{p{3.2cm}<{\centering} p{1cm}<{\centering}  p{2cm}<{\centering}}
\hline \hline

quantity & symbol & magnitude \\ [0.5ex] \hline 

lattice constant & $a$ & nm \\

$s-d$ exchange & $\Gamma$ & 0.1eV \\

$s-d$ exchange time & $\tau_{ex}$ & $10^{-14}$s \\

spin relaxation time & $\tau_{sf}$ & $10^{-12}$s \\

spin diffusion length & $\lambda_{N}$ & $10$nm \\

spin density & $n_{0}$ & $10^{27}/$m$^{3}$ \\


spin Hall angle & $\theta_{H}$ & $0.1$ \\

intrinsic DM & $D_{\alpha}^{0}$ & $10^{-3}$eV \\

superexchange & $J$ & $0.1$eV \\

Eq.~(\ref{DM_E_field}) & ${\overline w}$ & $10^{-19}$C \\

electric flux quantum & $\tilde{\Phi}_{E}^{0}$ & 1V \\




\hline 
\hline
\end{tabular}	
\caption{List of material parameters and their order of magnitude values. }
\label{tbl:parameter_values}
\end{table}

{\it Spin ejection and delivery of electricity.-} We now address the spin ejection from the CMI to an attached normal metal (NM). A spin current is induced in the NM when a localized spin ${\bf S}_{i}$ at the NM/CMI interface rotates \cite{Zhang04,Kajiwara10}. Defining the conduction electron spin ${\bf m}({\bf r},t)=-\langle{\boldsymbol\sigma}\rangle/2$, the $s$-$d$ coupling at the interface $H_{sd}=\Gamma{\boldsymbol\sigma}\cdot{\bf S}_{i}$ defines a time scale $\tau_{ex}=\hbar/2S|{\Gamma}|$, with $\Gamma<0$  \cite{Zhang04}. The Bloch equation in the NM reads 
\begin{eqnarray}
\frac{\partial{\bf m}}{\partial t}+{\boldsymbol\nabla}\cdot{\cal J}_{s}=\frac{1}{\tau_{ex}}{\bf m}\times{\hat{\bf S}}_{i}-\frac{\delta{\bf m}}{\tau_{sf}}
\label{Bloch_eq}
\end{eqnarray}
where ${\cal J}_{s}={\bf J}_{s}^{NM}\varotimes{\boldsymbol\sigma}\hbar/2$ is the spin current tensor, and $\tau_{sf}$ is the spin relaxation time in the NM. In equilibrium, we assume ${\bf m}$ hybridizes with each ${\bf S}_{i}$ on the spiral texture locally. If the dynamics of ${\bf S}_{i}$ is slow compared to $1/\tau_{ex}$, which is true for the proposed mechanism and also for other usual means such as ferromagnetic resonance\cite{Kajiwara10}, ${\bf m}$ follows $-{\hat{\bf S}}_{i}$ at any time with a very small deviation ${\bf m}={\bf m}_{0}+\delta{\bf m}=-n_{0}{\hat{\bf S}}_{i}+\delta{\bf m}$, where $n_{0}$ is the local equilibrium spin density. The spin current tensor ${\cal J}_{s}=-D_{0}{\boldsymbol\nabla}\delta{\bf m}$ is obtained from the diffusion of $\delta{\bf m}$, where $D_{0}$ is the spin diffusion constant. Under such an adiabatic process, the small deviation is\cite{Zhang04} 
\begin{eqnarray}
\delta{\bf m}=\frac{\tau_{ex}}{1+\xi^{2}}\left\{-\xi n_{0}\frac{\partial{\hat{\bf S}}_{i}}{\partial t}-n_{0}{\hat{\bf S}}_{i}\times\frac{\partial{\hat{\bf S}}_{i}}{\partial t}\right\}\;,
\label{dm_time}
\end{eqnarray}
where $\xi=\tau_{ex}/\tau_{sf}<1$ so one can drop the first term on the right hand side, and replace ${\hat{\bf S}}_{i}\times\partial_{t}{\hat{\bf S}}_{i}\rightarrow \delta(r){\hat{\bf S}}_{i}\times\partial_{t}{\hat{\bf S}}_{i}$ since ${\hat{\bf S}}_{i}$ is located at the NM/CMI interface $r=0$ ($ r$ as coordinate perpendicular to the interface). The resulting equation solves the time dependence of $\delta{\bf m}$. Away from $r=0$, Eq.~(\ref{Bloch_eq}) yields $D_{0}\nabla^{2}\delta{\bf m}=\delta{\bf m}/\tau_{sf}$, which solves the spatial dependence of $\delta{\bf m}$. The spin current caused by a particular ${\bf S}_{i}$ then follows 
\begin{eqnarray}
J_{s}^{NM}\delta{\hat{\bf m}}=\delta{\bf m}\frac{D_{0}}{\lambda_{N}}=-\frac{\tau_{ex}n_{0}D_{0}}{\left(1+\xi^{2}\right)\lambda_{N}}{\hat{\bf S}}_{i}\times\frac{\partial{\hat{\bf S}}_{i}}{\partial t}e^{-r/\lambda_{N}},
\label{ejected_spin_current}
\end{eqnarray} 
where $\lambda_{N}=\sqrt{D_{0}\tau_{sf}}$, similar to results obtained previously \cite{Kajiwara10}. 
If only the in-plane Goldstone mode is excited, as shown in Fig.~\ref{fig:E_field_spiral}, it is equivalent to a global rotation of spins ${\hat{\bf S}}_{i}=(-1)^{i}(\sin(\theta(t)+{\bf Q}\cdot{\bf r}_{i}),0,\cos(\theta(t)+{\bf Q}\cdot{\bf r}_{i}))$ in the field-free region. Thus the time dependence in Eq.~(\ref{ejected_spin_current}), ${\hat{\bf S}_{i}}\times\partial_{t}{\hat{\bf S}_{i}}={\hat {\bf y}}\partial\theta/\partial t$, is that described by Eq.~(\ref{dtheta_dt}), and is the same for every ${\bf S}_{i}$ at the NM/CMI interface, even though each ${\bf S}_{i}$ point at a different polar angle. In other words, the spin current ejected from each ${\bf S}_{i}$ of the AF spiral, described by Eq.~(\ref{ejected_spin_current}), is the same, so a uniform spin current flows into the NM.

We propose two setups to convert the ejected spin current into an electric signal. The first device uses inverse SHE \cite{Kajiwara10} in a NM deposited at the side of the spiral plane, yielding $\delta{\hat{\bf m}}$ perpendicular to ${\bf J}_{s}^{NM}$ and consequently a voltage in the transverse direction, as shown in Fig.~\ref{fig:E_field_spiral} (a). The second design ejects spin into a NM film deposited on top of the spiral plane, as shown in Fig.~\ref{fig:E_field_spiral} (b), causing $\delta{\hat{\bf m}}$ parallel to ${\bf J}_{s}^{NM}$. A spin accumulation in the NM develops and oscillates with time, producing an oscillating magnetic flux $\Phi_{B}$ through a coil that wraps around the NM, hence a voltage ${\cal E}=-\partial\Phi_{B}/\partial t$.

{\it Experimental realizations.-} The Raman scattering data on BFO \cite{Rovillain10} show that applying $|{\bf E}|\sim 100$kV/cm can change the spin wave velocity by $\delta v_{0}/v_{0}\sim 1\%$. We can make use of this information to estimate the field-dependence $\overline{w}$ in Eq.~(\ref{DM_E_field}). The $\omega_{\bf k}^{-}$ mode in Eq.~(\ref{AF_spiral_eigen_modes}) near ${\bf k}=(0,0)$ is
\begin{eqnarray}
\omega_{\bf k\rightarrow 0}^{-}&=&2\sqrt{2}SJka\left[1+\frac{5}{16}\left(\frac{D_{a}^{2}k_{a}^{2}+D_{c}^{2}k_{c}^{2}}{J^{2}k^{2}}\right)\right]
\nonumber \\
&=&\left(v_{0}+\delta v_{0}\right)k\;,
\end{eqnarray}
where $v_{0}=2\sqrt{2}SJa$ is the spin wave velocity in the absence of DM interaction. Assuming $D_{a}\neq 0$, $D_{c}=0$, and ${\bf E}\perp{\bf a}$, the Raman scattering data gives ${\overline w}\sim 10^{-19}$C$\sim |e|$. We remark that a coplanar magnetic order can be mapped into a spin superfluid
 \cite{Halperin69,Chandra90} $\psi_{i}$ by 
\begin{eqnarray}
\langle{\bf S}_{i}\rangle=S\left(\sin\theta_{i},0,\cos\theta_{i}\right)=\sqrt{v}\left({\rm Im}\psi_{i},0,{\rm Re}\psi_{i}\right)\;,
\end{eqnarray}
where $v$ is the volume of the 3D unit cell. Within this formalism, the ${\bf E}$ field can induce quantum interference of the spin superfluid via magnetoelectric effect, in which the electric flux vector ${\boldsymbol\Phi}_{E}=\oint{\bf E}\times d{\bf l}$
is quantized \cite{Chen13_2,Chen14}. The flux quantum is $\tilde{\Phi}_{E}^{0}=2\pi J/\overline{w}$, which is $\tilde{\Phi}_{E}^{0}\sim 1$V for BFO, close to that ($\sim 10$V) obtained from current-voltage characteristics of a spin field-effect transistor \cite{Chen13_2}, indicating that strong spin-orbit interaction reduces the flux quantum to an experimentally accessible regime. For instance, BFO has a spiral wave length $2\pi/Q\sim 100$nm, so in a BFO ring of $\mu$m size, the number of spirals at zero field is $n_{Q}\sim 10$, and applying $|{\bf E}|\sim 1$kV/cm can change $n_{Q}$ by 1. Besides changing the winding number, we remark that the magnetoelectric
effect can also be used to affect the topological properties of a magnet in a different respect\cite{You14}. Table I lists the parameters and their order of magnitude values by assuming CMI has similar material properties as other
magnetic oxide insulators such as YIG, and we adopt lattice constant $a\sim 1$nm for both CMI and the NM for simplicity.

For the device in Fig.~\ref{fig:E_field_spiral}, consider the field $|{\bf E}^{0}|\sim 100$kV$/$cm oscillating with a household frequency $\omega\sim 100$Hz is applied to a range $L\sim 1$mm. This region covers $N=L/a\sim 10^{6}$ sites with a number of spirals $n_{Q}\sim 10^{4}$ at zero field. The ${\bf E}$ field changes the number of spirals to $n_{Q}\sim 10^{5}$ within time period $1/\omega\sim 0.01$s, so the spins at the boundary ${\bf S}_{N}$ wind with angular speed $\partial_{t}\theta_{N}\sim 10^{7}\sin\omega t$ which is enhanced by 5 orders of magnitude from the driving frequency $\omega$. To estimate the ejected spin current in Eq.~(\ref{ejected_spin_current}), we use the typical spin relaxation time $\tau_{sf}\sim 10^{-12}$s and length $\lambda_{N}\sim 10$nm for heavy metals \cite{Kajiwara10}. The $s$-$d$ coupling can range between \cite{Kajiwara10} $0.01$eV to $1$eV. We choose $\Gamma\sim 0.1$eV, which gives $\tau_{ex}\sim 10^{-14}$s. The spin Hall angle $\theta_{H}\sim 0.1$ has been achieved \cite{Seki08,Pai12}. To estimate $n_{0}$, we use the fact that the $s$-$d$ hybridization $\Gamma{\boldsymbol \sigma}\cdot{\bf S}_{i}$ is equivalent to applying a magnetic field ${\bf H}=2\Gamma{\bf S}_{i}/\mu_{0} g\mu_{B}$ locally at the interface atomic layer of the NM. Given the typical molar susceptibility $\chi_{m}\sim 10^{-4}$cm$^{3}/$mol and molar volume $V_{m}\sim 10$cm$^{3}/$mol, the interface magnetization of the NM is $n_{0}\mu_{B}=\chi_{m}{\bf H}/V_{m}\sim 10^{4}$C/sm, thus $n_{0}\sim 10^{27}/$m$^{3}$. The oscillating ${\bf E}$ field gives ${\hat{\bf S}_{i}}\times\partial_{t}{\hat{\bf S}_{i}}=\partial_{t}\theta_{N}{\hat {\bf y}}\sim {\hat{\bf y}}10^{7}$Hz, so the ejected spin current is $J_{s}^{NM}\sim 10^{24}\hbar/$m$^{2}$s. Using the design in in Fig.~\ref{fig:E_field_spiral}(a) to convert $J_{s}^{NM}$ into a charge current via inverse SHE yields $J_{c}^{NM}\sim 10^{4}$A/m$^{2}$, hence a voltage $\sim \mu$V oscillating with $\omega$ in a mm-wide sample. To use the setup in Fig.~\ref{fig:E_field_spiral}(b), a NM film of area $\sim$ 1 mm$^{2}$ and thickness $\sim 10$nm yields ${\cal E}\sim$mV oscillating with $\omega$.

In summary, we propose that for multiferroics that have coplanar AF spiral order, such as BFO, applying an oscillating ${\bf E}$ field with frequency as low as household frequency generates a coherent planar rotation of the spin texture whose frequency is many orders of magnitude enhanced. This coherent rotation activates the Goldstone mode of multiferroic insulators that remains unaffected by the crystalline anisotropy. The Goldstone mode can be used to deliver electricity at room temperature up to the extensions of magnetic domains, in a way that is free from the energy loss due to Gilbert damping if the sample is free from impurities. The needlessness of ${\bf B}$ field greatly reduces the energy consumption and increases the scalability of the proposed device, pointing to its applications in a wide range of length scales.

We thank exclusively P. Horsch, J. Sinova, H. Nakamura, Y. Tserkovnyak, D. Manske, M. Mori, C. Ulrich, J. Seidel, and M. Kl\"{a}ui for stimulating discussions.

\begin{eqnarray}
\hline\nonumber 
\end{eqnarray}
{\bf{\large Supplementary material}}
\begin{eqnarray}
\hline\nonumber 
\end{eqnarray}

{\bf I. Crystalline anisotropy in multiferroics}

\vspace{0.2cm}

First we demonstrate that because the wave length of the spiral order in multiferroics is typically $1\sim 2$ orders longer than the lattice constant, and the exchange coupling is typically few orders larger than the crystalline anisotropy energy, the spiral order remains truly incommensurate and very weakly affected by the crystalline anisotropy. For simplicity, we consider a spiral state with wave vector ${\bf Q}\parallel (1,0)$ and translationally invariant perpendicular to $ {\bf Q} $ such that the geometry can be reduced to a 1D problem. The classical elastic energy for a 1D antiferromagnetic (AF) spiral is 
\begin{eqnarray}
E_{0}&=&\sum_{n}-\tilde{J}_{a}S^{2}\cos\left(\theta_{n+1}-\theta_{n}-\theta_{a}\right)
\nonumber \\
&\approx&-N\tilde{J}_{a}S^{2}+\sum_{n}\frac{1}{2}\tilde{J}_{a}S^{2}\left(\theta_{n+1}-\theta_{n}-\theta_{a}\right)^{2}\;,
\end{eqnarray}
where $\theta_{n}={\bf Q}\cdot{\bf r}_{n}$ is the angle relative to the staggered spin $(-1)^{i}{\bf S}_{i}$, and $ \theta_{a}={\bf Q}\cdot{\bf a}$ is the natural pitch angle between neighboring spins ($ {\bf a} = (a,0)$). The square lattice symmetry of our model yields a $4$-fold degenerate crystalline spin anisotropy\cite{Sonin10}, leading to the total energy 
\begin{eqnarray}
E=\sum_{n}\frac{1}{2}\tilde{J}_{a}S^{2}\left(\theta_{n+1}-\theta_{n}-\theta_{a}\right)^{2}
\nonumber\\
+\sum_{n}V_{ani}\left(1-\cos 4\theta_{n}\right)\;,
\label{E_total}
\end{eqnarray} 
where $V_{ani}$ is the anisotropy energy per site. This is the well-known Frenkel-Kontorowa(FK) model\cite{Frenkel38,Frank49} that has been discussed extensively owing to its rich physics.

\begin{figure}[ht]
\begin{center}
\includegraphics[clip=true,width=0.95\columnwidth]{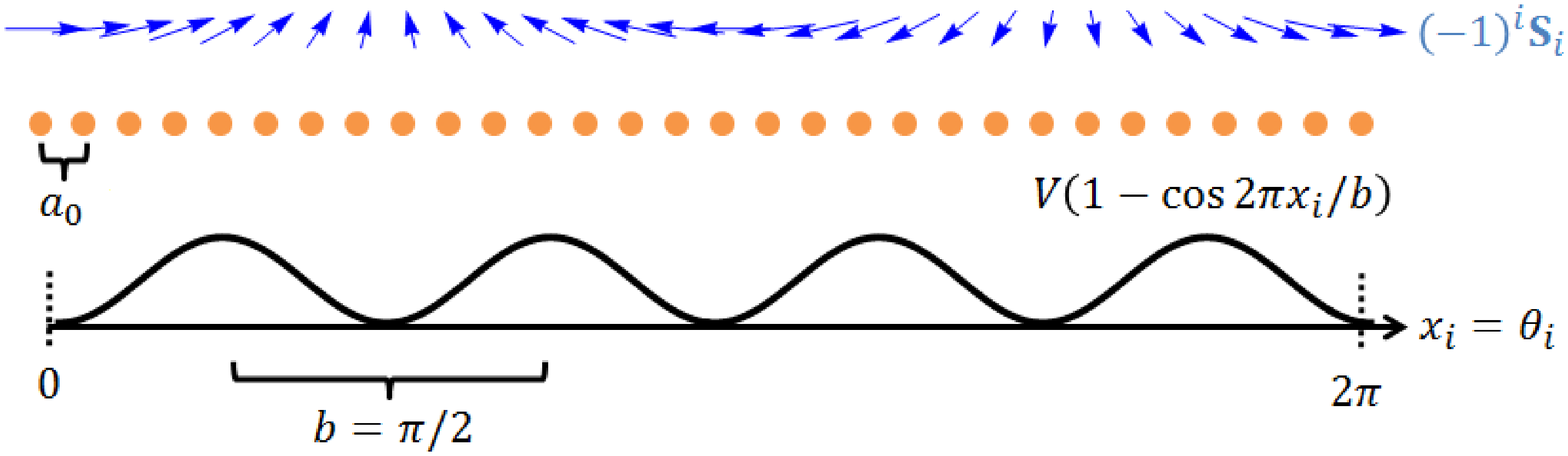}
\caption{ (color online) Schematics of mapping the AF spiral order in the presence of crystalline anisotropy into FK model. The angles $\theta_{i}$ of staggered spins $(-1)^{i}{\bf S}_{i}$ (blue arrows) are mapped into displacements $x_{i}$ of particles (orange dots). The width of the 4-fold degenerate pinning potential $V\left(1-\cos 2\pi x_{i}/b\right)$ is $b=\pi/2$, and the spacing of particles in the absence of the pinning potential is $a_{0}={\bf Q}\cdot{\bf a}$.  } 
\label{fig:mapping_to_FK}
\end{center}
\end{figure}

We consider the limit of weak anisotropy $ V = V_{ani} / \tilde{J}_a S^2 a^2 \ll 1 $ and the case of long wavelength of the spiral, $ \theta_a \ll \pi/2 $ 
where $ \pi / 2 $ is the angle between two minima of the anisotropy potential. In the spirit of Ref.\cite{Bak82,Theodorou78} we assume now that
there are prime numbers, $ M $ and $ L $ with $ M \tilde{\theta}_a = L \pi /2 $ and $ M \gg L $ which is the average pitch in the ground state of Eq.(\ref{E_total}). Then we introduce the parametrization
\begin{equation}
\theta_n = n \tilde{\theta}_a + \frac{\varphi_n}{4} 
\end{equation}
and the misfit parameter $ \delta = 4 (\theta_a - \tilde{\theta}_a) $. Turning to the continuous limit one can derive the effective energy functional based on expanding the first harmonic approximation\cite{Bak82,Theodorou78,Pokrovskii78},
\begin{equation}
\tilde{E}[\varphi] = \int dx \left[ \frac{1}{2} \left( \frac{d \varphi}{dx} - \delta \right)^2 + V_M \cos ( M \varphi) \right]
\end{equation}
with $ V_M \sim V^M $ which can become extremely small for $ M \gg 1 $. The commensurate-incommensurate transition happens if $ \delta $ is large enough to stabilize the formation of solitons $ \delta > \delta_c(M) \sim 4 \sqrt{V_M}/\pi $. Deep inside the incommensurate phase, $ \varphi (x) \approx \delta x $ such that $ \theta_n \approx \theta_a n $ follows esentially the natural spiral pitch. 

In our system, BFO, the spiral wave length $ \ell \approx 60 {\rm nm} \sim 100 a $ which yields $ M \sim 100/4 = 25 $, i.e. every 25$^{th}$ spin could be pinned along one of the 4 anisotropy minima (assuming $ L= 1 $). Typical anisotropy energies for ferrites\cite{Yosida57} lead to $ V_{ani} \sim 10^{-3} {\rm eV} $ while the exchange energy is $ J_a \sim 0.1 {\rm eV} $, from which we obtain $ V \sim V_{ani} / J_a \sim 10^{-2} $ and consequently $V_{M}\sim 10^{-50}$ is a negligible number. The misfit parameter may be as large as $ \delta = 4 (\theta_a - \tilde{\theta}_a) \sim \pi / M^2 $ such that $ \delta \gg \delta_c(M) $ is well satisfied, even if by an electrical field $ M $ shrinks by one order of magnitude. Thus, the electric field-driven oscillations of the spin spiral remains most likely unaffected by the spin anisotropy. The small $V_{M}$ renders the energy gap due to the anisotropy energy irrelevant, hence the in-plane magnon mode remains essentially undamped. Another important consequence of this analysis is that although the concept of spin superfluidity, i.e., treating the spin texture as a quantum condensate, has been proposed long ago, its realization in collinear magnets is problematic because of the crystalline anisotropy and subsequently the formation of domain walls. We demonstrate explicitly that multiferroics are not subject to these problems because of the noncollinear order, hence a room temperature macroscopic condensate of mm size can be realized.

\vspace{0.2cm}
{\bf II. Phase-pinning impurities in multiferroics}
\vspace{0.2cm}

We proceed to show that dilute, randomly distributed impurities, exist either in the bulk of the multiferroic or at the metal/multiferroic interface, do not obstruct the proposed electrically controlled multiferroic magnonics. Drawing analogy from the FK model, impurities that pin the spins along certain crystalline directions, denoted by phase-pinning impurities, are the impurities to be considered because they tend to impede the coherent motion of spins\cite{Chaikin95}. Since we propose to use an oscillating ${\bf E}$ field to drive the spin rotation from the boundary, each cross section channel is equivalent, which reduces the problem from 2D to 1D. This leads us to consider the following 1D classical model similar to Eq.~(\ref{E_total}) for the field-free region (${\bf S}_{N+1}^{\prime}$ to ${\bf S}_{N+M}^{\prime}$ in the Fig.~2 of the main text).
\begin{eqnarray}
E=\sum_{i}\frac{1}{2}\tilde{J}_{a}S^{2}\left(\theta_{i+a}-\theta_{i}-\theta_{a}\right)^{2}
-\sum_{i\in imp}V_{imp}\cos 4\theta_{i}\;,
\label{E_total_imp}
\end{eqnarray} 
where $V_{imp}>0$ is the pinning potential, and $\sum_{i\in imp}$ sums over impurity sites. The total length of the chain is $L^{\prime}=Ma$ with $M$ an integer. In the presence of oscillating ${\bf E}$ field that causes the winding of boundary spins (${\bf S}_{N}^{\prime}$ in the Fig.~2 of the main text), the angle of spins in the disordered field-free region has three contributions
\begin{eqnarray}
\theta_{i}=\theta_{i}^{0}+\Delta\theta_{i}+\eta_{i}\;,
\end{eqnarray}
where $\theta_{i}^{0}$ represents the spiral texture in the unstretched clean limit satisfying $\theta_{i+a}^{0}-\theta_{i}^{0}-\theta_{a}=0$, $\Delta\theta_{i}$ is the stretching of the spin texture caused by winding of boundary spins, and $\eta_{i}$ is the distortion due to impurities. Only the later two contribute to the elastic energy, so Eq.~(\ref{E_total_imp}) becomes
\begin{eqnarray}
E&=&\sum_{i}\frac{1}{2}\tilde{J}_{a}S^{2}\left(\Delta\theta_{i+a}-\Delta\theta_{i}+\eta_{i+a}-\eta_{i}\right)^{2}
\nonumber\\
&-&\sum_{i\in imp}V_{imp}\cos 4\theta_{i}\;.
\label{E_total_imp_dtheta}
\end{eqnarray} 
In this analysis we consider weak impurities $V_{imp}\ll\tilde{J}_{a}S^{2}$, and assume that the winding of the boundary spins is slow such that the winding spreads through the whole field-free region evenly, causing every pair of neighboring spins to stretch by the same amount $\Delta\theta_{i+a}-\Delta\theta_{i}=\Delta\theta$. For the electrically driven magnonics proposed in the main text, which can achieve winding of boundary spins by $\theta_{N}\sim n_{Q}\sim 10^{5}$ within half-period, a field-free region of length $L^{\prime}\sim$mm has $\Delta\theta\sim 0.1$, so our numerics is done with $\Delta\theta$ limited within this value.


In the weak impurity limit, the length scale $L_{0}$ over which $\eta_{i}$ changes by ${\cal O}(1)$ can be calculated in the following way. The elastic energy part in Eq.~(\ref{E_total_imp_dtheta}) within $L_{0}$ is, in the continuous limit,
\begin{eqnarray}
K(L_{0})&=&\frac{1}{a}\int_{0}^{L_{0}}dx\frac{1}{2}\tilde{J}_{a}S^{2}a^{2}\left(\frac{\Delta\theta}{a}+\partial_{x}\eta\right)^{2}
\nonumber\\
&=&\frac{L_{0}}{2a}\tilde{J}_{a}S^{2}\Delta\theta^{2}+\frac{\tilde{J}_{a}S^{2}\Delta\theta}{\alpha_{1}}+\frac{\tilde{J}_{a}S^{2}a}{2\alpha_{0}L_{0}}\;,
\label{FLR_elastic_energy}
\end{eqnarray}
where $\alpha_{0}$ and $\alpha_{1}$ are numerical constants of ${\cal O}(1)$, and are set to be unity without loss of generality. Denoting impurity density as $n_{imp}=N_{imp}/L^{\prime}$ where $N_{imp}$ is the total number of impurities in the sample, the impurity potential energy within $L_{0}$ is calculated by
\begin{eqnarray}
V(L_{0})&=&-V_{imp}{\rm Re}\left(\sum_{i\in imp}e^{4i\left(\theta_{i}^{0}+\Delta\theta+\overline{\eta}\right)}\right)
\nonumber\\
&=&-V_{imp}\sqrt{n_{imp}L_{0}}\;.
\label{VL0_fluctuation}
\end{eqnarray}
Note that the contribution comes not from the zeroth order impurity averaging, but its fluctuation that mimics a random walk in the complex plane\cite{Fukuyama78}. The phase $\overline{\eta}$ is assumed to be constant within $L_{0}$ and chosen to give Eq.~(\ref{VL0_fluctuation}) and hence the total energy $E(L_{0})=K(L_{0})+V(L_{0})$ within $L_{0}$. Minimizing the total energy per site $E(L_{0})/L_{0}$ gives the most probable pinning length $L_{0}$. In the unstretched case $\Delta\theta=0$,
\begin{eqnarray}
L_{0}=\left(\frac{\tilde{J}_{a}S^{2}a}{\alpha_{0}V_{imp}n_{imp}^{1/2}}\right)^{2/3}
\label{FLR_length_unstretched}
\end{eqnarray}
is similar to the Fukuyama-Lee-Rice (FLR) length that characterizes the impurity pinning of a charge density wave ground state\cite{Fukuyama78,Lee79}. Putting Eq.~(\ref{FLR_length_unstretched}) back to Eqs.~(\ref{VL0_fluctuation}) and (\ref{FLR_elastic_energy}), the corresponding $E(L_{0})a/L_{0}<0$ can be viewed as the pinning energy per site that impedes the coherent rotation of spins, and equivalently represents the gap opened at the Goldstone mode.

\begin{figure}[ht]
\begin{center}
\includegraphics[clip=true,width=0.95\columnwidth]{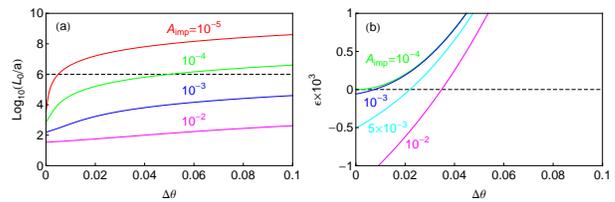}
\caption{ (color online) (a) The logrithmic of the dimensionless FLR length $\log_{10}\left(L_{0}/a\right)$ versus winding angle per site $\Delta\theta$, in several values of the dirtiness parameter $A_{imp}$. Dashed line indicates the threshold when $L_{0}\sim$mm. (b) The dimensionless pinning energy $\epsilon$ versus winding per site.  } 
\label{fig:FLR_length_E0}
\end{center}
\end{figure}

In the presence of the stretching $\Delta\theta$, the expression of $L_{0}$ is rather lengthy. It is convenient to define two dimensionless parameters 
\begin{eqnarray}
A_{imp}&=&\frac{V_{imp}}{\tilde{J}_{a}S^{2}}\sqrt{N_{imp}\frac{a}{L^{\prime}}}\;,
\nonumber \\
\epsilon&=&\frac{E(L_{0})}{\tilde{J}_{a}S^{2}}\left(\frac{a}{L_{0}}\right)\;,
\end{eqnarray}
where $A_{imp}$ (the "dirtiness parameter") is the impurity potential measured in unit of the elastic constant times the square root of the impurity density, and $\epsilon$ is the total energy per site measured in unit of the elastic constant. Figure \ref{fig:FLR_length_E0} shows the logrithmic of the dimensionless pinning length $L_{0}/a$ and the dimensionless total energy $\epsilon$, plotted as functions of the stretching $\Delta\theta$. There are two evidences showing that the spin texture, originally pinned by impurities with the pinning length in Eq.~(\ref{FLR_length_unstretched}), is depinned by the stretching $\Delta\theta$: Firstly, the pinning length $L_{0}$ increases as increasing $\Delta\theta$. For a particular sample size, for instance $L^{\prime}\sim$mm, the spin texture is depinned when the pinning length exceeds the sample size $L_{0}> L^{\prime}$, or equivalently when $\Delta\theta$ is greater than a certain threshold (intercept of the dashed line and the colored lines in Fig.~\ref{fig:FLR_length_E0} (a)). Secondly, the pinning energy $\epsilon$ becomes positive at large $\Delta\theta$, indicating that the elastic energy from stretching overcomes the impurity pinning energy, so the spin texture is depinned. From Fig.~\ref{fig:FLR_length_E0}, it is also evident that the cleaner is the sample, the easier it is to depin the spins by stretching, as smaller $A_{imp}$ requires smaller threshold value of $\Delta\theta$. We conclude that the phase pinning impurities do not hamper the proposed electrically driven multiferroic magnonics as long as the dirtiness of the sample is limited, the winding speed of the boundary spin is sufficient, and the sample size is short enough.

\end{document}